\documentclass[reprint,superscriptaddress,amsmath,amssymb,aps,prl,longbibliography]{revtex4-2}
\usepackage[utf8]{inputenc}
\usepackage{graphicx}
\usepackage{dcolumn}
\usepackage{bm}
\usepackage[colorlinks,citecolor=blue,linkcolor=blue,urlcolor=blue]{hyperref}
\usepackage[normalem]{ulem}

\begin{document}

\title{Enhancement of $d$-wave Pairing in Strongly Correlated Altermagnet}

\author{Jianyu Li}
\thanks{These authors contributed equally.}
\affiliation{School of Science, Harbin Institute of Technology, Shenzhen, 518055, China}
\affiliation{Shenzhen Key Laboratory of Advanced Functional Carbon Materials Research and Comprehensive Application, Shenzhen 518055, China.}

\author{Ji Liu}
\thanks{These authors contributed equally.}
\affiliation{School of Science, Harbin Institute of Technology, Shenzhen, 518055, China}
\affiliation{Shenzhen Key Laboratory of Advanced Functional Carbon Materials Research and Comprehensive Application, Shenzhen 518055, China.}

\author{Xiaosen Yang}
\email{yangxs@ujs.edu.cn}
\affiliation{Department of Physics, Jiangsu University, Zhenjiang, 212013, China.}

\author{Ho-Kin Tang}
\email{denghaojian@hit.edu.cn}
\affiliation{School of Science, Harbin Institute of Technology, Shenzhen, 518055, China}
\affiliation{Shenzhen Key Laboratory of Advanced Functional Carbon Materials Research and Comprehensive Application, Shenzhen 518055, China.}

\date{\today}

\begin{abstract} 
Altermagnetism, featuring momentum-dependent spin splitting without net magnetization, has attracted a growing interest for spintronics. We study a Fermi Hubbard model with altermagnetic order arising from the spin-anisotropic hopping near half-filling using constrained-path quantum Monte Carlo. Spin-dependent hopping breaks SU(2) symmetry and disrupts Fermi surface nesting, giving rise to an altermagnetic state with momentum-space spin splitting but no net magnetization. We find that increasing anisotropy suppresses long-range antiferromagnetic order and significantly enhances effective $d$-wave pairing correlations. Our results demonstrate a doping-free route to unconventional superconductivity mediated by short-range spin fluctuations in an altermagnetic background.

\end{abstract} 
\maketitle

\emph{Introduction.}-- Altermagnetism (AM) is a novel magnetic state combining features of ferromagnetism (FM) and antiferromagnetism (AFM), exhibiting zero net magnetization and momentum-dependent spin splitting through real-space rotational symmetry~\cite{Smejkal2022-wp, mazinEditorialAltermagnetismNew2022a, aminNanoscaleImagingControl2024a, baiAltermagnetismExploringNew2024a, jiangEnumerationSpinspaceGroups2024a, mcclartyLandauTheoryAltermagnetism2024a, rimmlerNoncollinearAntiferromagneticSpintronics2024a, jungwirthAltermagnetismUnconventionalSpinordered2025a, songAltermagnetsNewClass2025a, zhangCrystalsymmetrypairedSpinValley2025}. It enables spin-split bands without spin-orbit coupling, offering advantages for spintronics\cite{fangQuantumGeometryInduced2024a, krempaskyAltermagneticLiftingKramers2024a, reimersDirectObservationAltermagnetic2024a,xuFrustratedAltermagnetismCharge2023a,ZhangPhysRevLettPredictable,GuPhysRevLettFerroelectricSwitchable}. Due to the time-reversal breaking band structure and alternating momentum-dependent spin splitting, AM can enrich interesting effects and phases such as crystal Hall effect~\cite{doi:10.1126/sciadv.aaz8809, satoAltermagneticAnomalousHall2024a,reichlovaObservationSpontaneousAnomalous2024a,brekkeMinimalModelsTransport2024a,JinPhysRevLettSkyrmionHall}, giant tunneling magnetoresistance~\cite{Shao2021-ge,Smejkal2022-yw}, finite-momentum Cooper pairing~\cite{zhangFinitemomentumCooperPairing2024a, Cao2024-hl, ChakrabortyPhysRevBZerofield, hongUnconventionalPwaveFinitemomentum2025a, Cao2025-xt}, unconventional topological phases~\cite{antonenkoMirrorChernBands2025a, gonzalez-hernandezSpinChernNumber2025a, ghorashiAltermagneticRoutesMajorana2024a, LiPhysRevBhigherorder} and spin-momentum interactions~\cite{reichlova2024observation, linCoulombDragAltermagnets2024a, ouassouDcJosephsonEffect2023a}. Recent advances have significantly deepened our understanding of AM, uncovering its realization mechanisms across a broad range of physical systems. However, only a limited number of intrinsic altermagnetic materials have been experimentally confirmed to date~\cite{krempasky2024nature,fedchenko2024sciadv,lee2024prl,zeng2024advsci,reimers2024natcomm,osumi2024prb,liu2024prl,jiang2024arxiv,zhang2024arxiv}. 

Efforts to identify candidate materials have largely focused on tuning electronic structure via chemical and structural design, for example, anisotropic spin splitting has been attributed to orbital anisotropy in systems with symmetry-breaking local environments~\cite{brekke2023twodimensional,leeb2024spontaneous,giuli2025altermagnetism,ferrari2024altermagnetism,kaushal2024altermagnetism,antonenko2025mirror}. More recently, a variety of physically motivated design strategies have been proposed to engineer AM. Beyond small-spin-cluster engineering and symmetry-guided top-down approaches~\cite{Zhu2025-ci,PhysRevLett.134.176902}, researchers have explored tuning orbital hybridization through selective ligand coordination to create sublattice-dependent exchange pathways~\cite{naka2025altermagnetic,chu2025asymmetric}, applying epitaxial strain or external pressure to break inversion symmetry while preserving C$_2$/C$_4$ rotational symmetry~\cite{zhou2025manipulation,doi:10.1126/sciadv.adj4883}, and exploiting staggered crystal-field environments or single-ion anisotropies in low-symmetry lattices to generate anisotropic exchange interactions~\cite{yang2025natcomm,gonzalez2024anisotropic}.

\begin{figure}[b!]
	\centering
    \includegraphics[width=\linewidth]{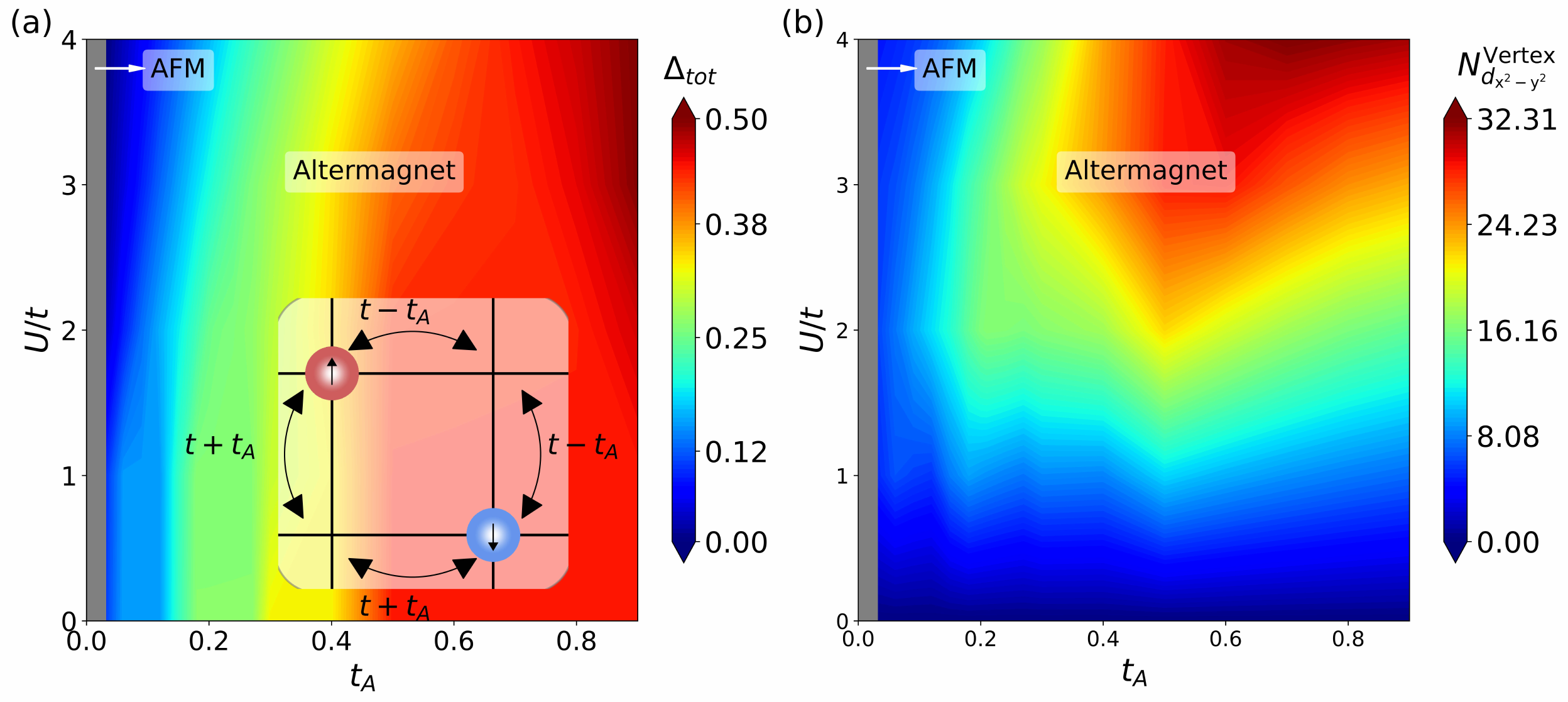}
    \caption{(Color online) (a) Total momentum domain spin polarization $\Delta_{tot}$ of varying interaction $U/t$ and anisotropy $t_A$, altermagnetism~(AM) with finite $\Delta _{tot}$ overtakes antiferromagnetism~(AFM) once the anisotropy on. (b) The total $d$-wave vertex pairing function $N^{\rm Vertex}_{\mathrm d_{x^2-y^2}}$ that enhanced by both interaction and anisotropy in altermagetic phase. The result is obtained using the quantum Monte Carlo method on $16 \times 16$ lattice close to half-filling.
    }\label{fig1}
\end{figure}

An important avenue toward understanding AM is to investigate the role of electronic interactions, particularly in narrow-band systems with sizable screened Coulomb repulsion~\cite{vsmejkal2022giant,maierWeakcouplingTheoryNeutron2023a,dasRealizingAltermagnetismFermiHubbard2024a,bose2024altermagnetism,ferrari2024altermagnetism,zhao2025altermagnetism,giuli2025altermagnetism,leeb2024spontaneous,das2024realizing,durrnagel2024arxiv,yu2024vanhove,roig2024arxiv}. AM has been linked to $d$-wave superconductivity and phase separation in multiorbital $t-J$ models~\cite{bose2024altermagnetism}, as well as to interaction-induced spin-charge conversion~\cite{giuli2025altermagnetism}. Orbital ordering offers a further route to AM in the absence of crystal anisotropy~\cite{leeb2024spontaneous}. Ultracold atomic systems provide a promising platform to explore these effects in a controllable setting~\cite{das2024realizing}. However, most prior studies rely on mean-field approximations, and the role of strong correlations remains largely unaddressed. This motivates the use of unbiased many-body approaches, such as quantum Monte Carlo, to uncover correlation-driven altermagnetic phases.

In the strongly correlated regime, a central issue is the interplay between magnetism and pairing. A comprehensive understanding of pairing mechanisms in altermagnetic systems with repulsive interactions remains an open and fundamentally important problem, especially as AM has recently been proposed as an interaction-induced instability in cuprates~\cite{Li2024-op}. Previous studies have explored superconducting states in AM metals, highlighting finite-momentum pairing driven by Fermi surface anisotropy~\cite{sim2024pair,zhang2024finite,chakraborty2024zero} and topological superconductivity arising from the lifted Kramers degeneracy~\cite{zhu2023topological,brekke2023twodimensional,banerjee2024altermagnetic}. Unconventional pairing states have been reported, including $d$-wave pairing and pair density wave orders in multiorbital $t-J$ models~\cite{Bose2024-gh}, and functional renormalization group studies have revealed competing $d$-wave singlet and pair density wave instabilities at finite temperature~\cite{Parthenios2025-ak}. These results suggest a rich competition between AM order and interaction-driven superconducting instabilities.

In this work, we study a minimal yet representative model: the Fermi Hubbard Hamiltonian with spin-dependent anisotropic hopping near half-filling. We show that introducing spin-dependent anisotropy induces an altermagnetic phase with momentum-dependent spin splitting. The anisotropy suppresses the repulsion-driven antiferromagnetic order and concurrently enhances unconventional $d$-wave~($d_{x^2-y^2}$-wave) pairing correlations in the strongly correlated regime. This enhancement arises from the interplay between the on-site Hubbard interaction and the anisotropic spin-dependent Fermi surface, which breaks the perfect nesting present at half-filling. We suggest a possible mechanism behind the phenomena by studying $t-J$ model with altermagnetic dispersion at half-filling.
\begin{figure}[t]
	\centering
    \includegraphics[width=\columnwidth]{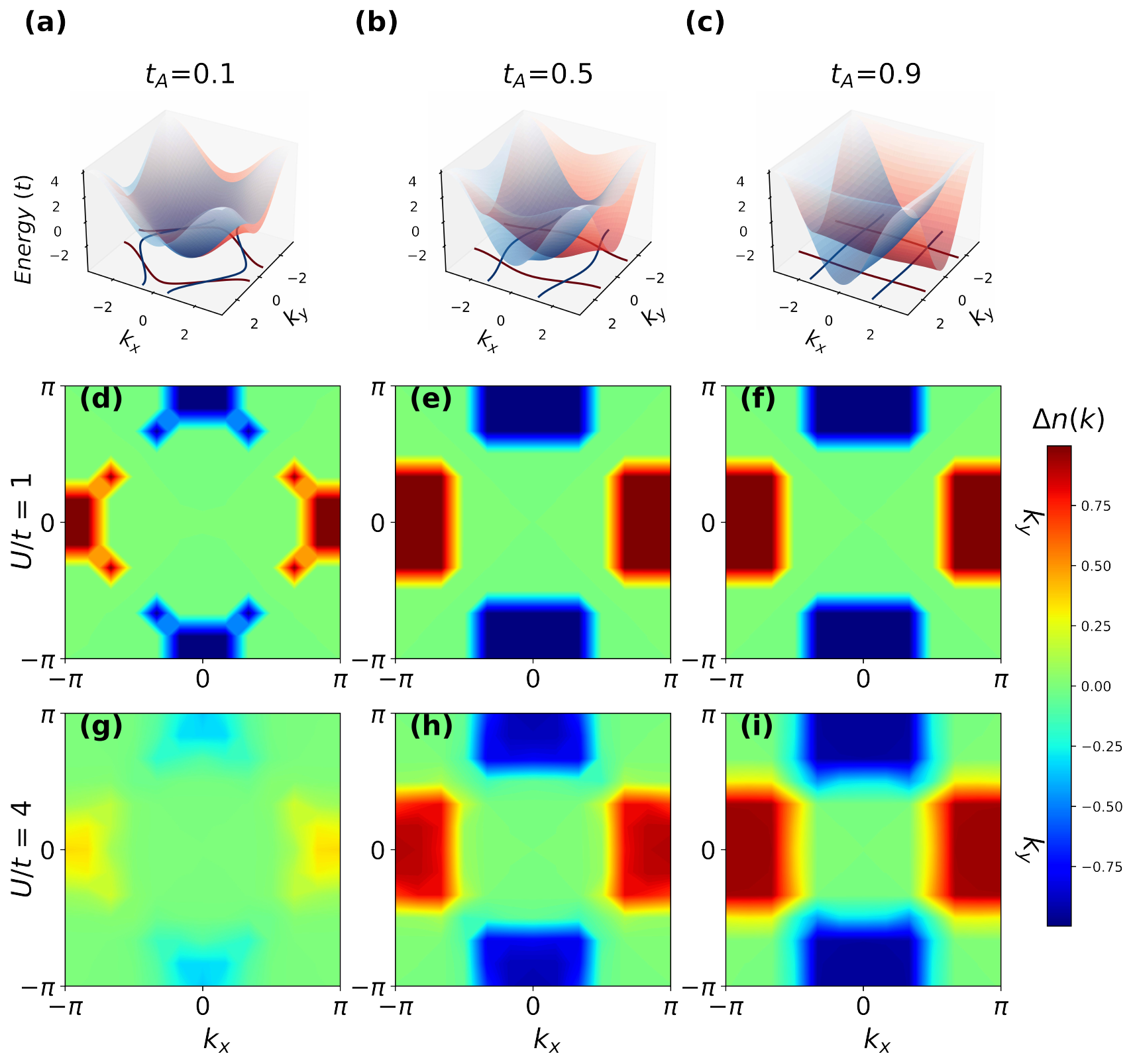}
    \caption{(Color online) Momentum-resolved spin polarization \( \Delta n(k) = n_{\uparrow}(k) - n_{\downarrow}(k) \). (a-c) Schematic diagrams of Fermi surface of up-spin~(red) and down-spin~(blue) with different $t_A$. (d-f) $\Delta n(k)$ for $U/t=1$ and (g-i) $U/t=4$. The suppression of $\Delta n(k)$ is observed when increasing $U/t$ from 1 to 4.}\label{fig2}    
\end{figure}

\emph{The altermagnetic Hubbard model}-- Here, we re-visit the Hamiltonian of Fermi Hubbard model on a square lattice with spin-dependent anisotropic hopping given by
\begin{equation}
H =-\sum_{\substack{i, \sigma, l=\hat{x},\hat{y}}} \left( t_{l,\sigma} c_{i,\sigma}^\dag c_{i+l,\sigma} + h.c. \right) + U \sum_i n_{i,\uparrow} n_{i,\downarrow},  
\label{eq:hamiltonian}
\end{equation}
with spin-dependent anisotropic hopping amplitudes $t_{l,\sigma}$, on-site interaction $U > 0$, $c_{i,\sigma}^\dag$ ($c_{i,\sigma}^{\,}$) creation (annihilation) operators with spin $\sigma$ = $\uparrow,\downarrow$, and  $n_{i,\sigma}=c_{i,\sigma}^\dag c_{i,\sigma}$. We give a schematic diagram on 2D square lattice with spin-dependent anisotropic hopping in the inset of Fig.~\ref{fig1}(a). To define the anisotropy of fermion hopping, we define a variable $t_A$, where $t_{\hat{y}\downarrow} = t_{\hat{x}\uparrow} = t-t_A$, $t_{\hat{x}\downarrow} = t_{\hat{y}\uparrow} = t + t_A$, leading to an unpolarized system with balanced spin populations, $\langle n_{i,\uparrow} \rangle = \langle n_{i,\downarrow} \rangle = n/2$. Without loss of generality, we assume nearest-neighbor hopping $t=1$ and the anisotropy parameter $t_A \in [0,1]$. Then, $t_A=0$ corresponds to the isotropic Hubbard model with $C_4$ symmetry and $t_A=1$ is the extreme anisotropic limit with $C_2$ symmetry, where fermions of one species can only hopping in one direction.

The spin splitting phenomenon is one of the most important characteristic of AM, in which we introduce a quantity, total momentum domain spin polarization $\Delta _{tot}$, to quantify it:
\begin{equation}
\Delta _{tot}=\sum_k{|\Delta n\left( k \right) |}=\sum_k{|n_{\uparrow}\left( k \right) -n_{\downarrow}\left( k \right) |},  
\label{eq:totalsp}
\end{equation}
where $\Delta n_{k}$ is the spin polarization in momentum space. The non-zero value of this quantity represents the emergence of altermagnetism in the model for zero net magnetization. We study the system using the constrained-path quantum Monte Carlo~(CPQMC) with a Hartree-Fock trial function. The method uses an constraint path approximation, and is widely used in the system with sign problem to uncover the ground states ~\cite{Zhang1995-hn, Zhang1997-ry, Shi2021-iu}. We have tested its convergence of different hyperparameters, and the details of the measurements are given in the supplementary materials.

\emph{Momentum-resolved spin polarization}-- We mainly investigate the properties of the ground states close to half filling, in which a perfect nesting of Fermi surface is achieved in the absence of spin-dependent anisotropy. This leads to the magnetic instability and result in an antiferromagnetic phase. The spin-dependent hopping anisotropy destroys the nesting, which might result into the competition of orders other than AFM phase. Fig.~\ref{fig1}(a) illustrates how total momentum domain spin polarization changes with the anisotropy parameter \(t_A\) and the Hubbard interaction strength \(U/t\). As \(U/t\) increases, the repulsion promoting electron localization and the formation of well-defined magnetic moments on lattice sites. This localization facilitates virtual hopping processes, yielding an effective antiferromagnetic coupling between neighboring spins. The total momentum domain spin polarization increases with \(t_A\), determined by reducing overlaps between the occupied momentum space of spin-\(\uparrow\) and spin-\(\downarrow\) fermions.

As increasing of \(t_A\), the electronic band structure of different spins becomes increasingly momentum-dependent, as shown in Fig.~\ref{fig2}(a–c). When Hubbard-type electron–electron interactions increase, comparing Fig.~\ref{fig2}(d–f) and Fig.~\ref{fig2}(g–i), it leads to increasing self-energy corrections that renormalize the quasiparticle dispersion by fermion scattering. These self-energy corrections lead to slight enlargement of momentum space with spin splitting, and reduced total momentum domain spin polarization. Our results confirm the persistency of altermagnetism under the Hubbard interaction at different levels of anisotropy. 

\begin{figure}[t!]
	\centering
    \includegraphics[width=\columnwidth]{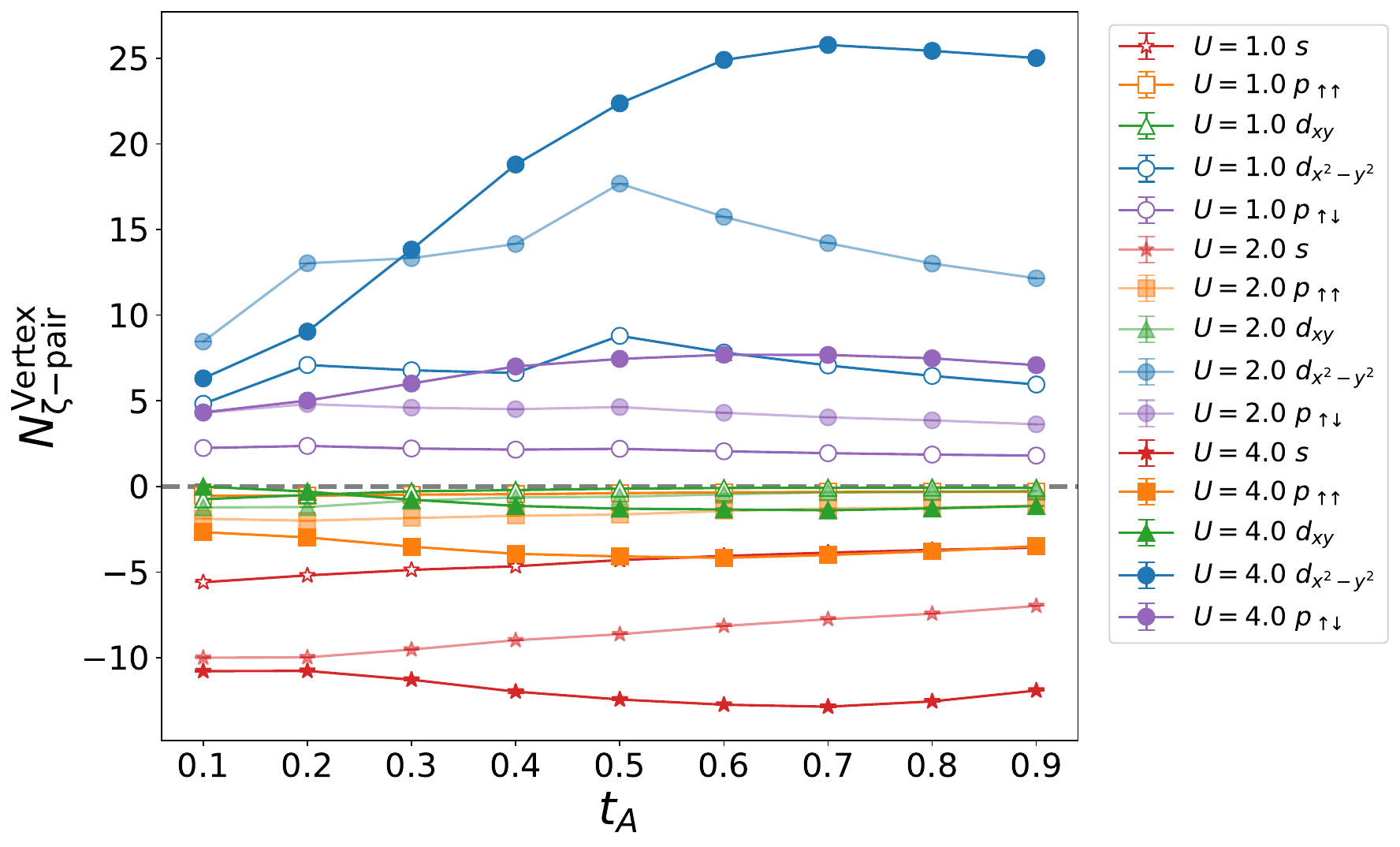}
    \caption{(Color online) The trend of the total vertex pairing function in different channels at $n \sim 1$ as a function of the anisotropy $t_A$, with $U=1$~(empty), $U=2$~(light color) and $U=4$~(deep color). We found enhancement of vertex pairing by Hubbard repulsion and anisotropy in $d$-wave and $p_{\uparrow,\downarrow}$-wave channels, while suppression are found in other channels.
    }\label{fig3}
\end{figure}

Previous theoretical works have demonstrated that the Hubbard interaction $U$ can enhance altermagnetic order in systems with anisotropic local orbitals, often via orbital-selective exchange mechanisms or orbital ordering in multiorbital frameworks~\cite{giuli2025altermagnetism,ferrari2024altermagnetism,ferrariAltermagnetismShastrySutherlandLattice2024a,das2024realizing}. In contrast, our results reveal a qualitatively different behavior in a single-orbital Hubbard model with spin-dependent hopping anisotropy. Specifically, we find that increasing $U$ leads to a suppression of the total momentum domain spin polarization $\Delta_{\text{tot}}$, a key signature of altermagnetism. 
The origin of this discrepancy lies in the model's orbital composition. Unlike multiorbital systems where Hubbard $U$ facilitates orbital polarization and interorbital exchange conducive to altermagnetic symmetry breaking, our model features only a single orbital per site. Consequently, the effect of $U$ manifests primarily through localization and incoherence, which compete with the spin-dependent kinetic anisotropy responsible for the altermagnetic band structure. This competition leads to a net reduction in the spin-split spectral weight, despite persistent anisotropy.

\emph{Enhancement of d-wave pairing}-- The vertex pairing function captures the vertex contribution from the interaction, where its calculation in momentum space is as following:
\begin{equation}
N^{\rm Vertex}_{\mathrm \zeta}({\bf k}) = (1/N)\sum_{i,j} \mbox{exp}[i{\bf k}({\bf r}_i-{\bf r}_j)]C^{\rm Vertex}_{\mathrm \zeta}(i,j),
\label{nspdtkpair}
\end{equation}
where $\zeta$ represents the pairing channel, the effective real-space $d$-wave pairing operator $C^{\rm Vertex}_{d_{x^2-y^2}}(i,j) = \langle {\Delta}_{d_{x^2-y^2}}^{\dagger}(i) {\Delta}_{d_{x^2-y^2}}(j) \rangle - \sum_{\delta_{\zeta},\delta'_{\zeta}}G^{\uparrow}_{i,j}G^{\downarrow}_{i+\delta_{\zeta},j+\delta'_{\zeta}}$, with $G^{\sigma}_{i,j}= \langle c_{i\sigma}c^\dagger_{j\sigma} \rangle$, $\Delta_{d_{x^2-y^2}}^\dagger(i)=c^\dagger_{i\uparrow}(c^\dagger_{i+x \downarrow}-c^\dagger_{i+y \downarrow}+c^\dagger_{i-x \downarrow}-c^\dagger_{i-y \downarrow})$, and  $\delta^{(')}_{\zeta}$ as the NN sites. For other pairing channels, we give the detailed definition in the supplementary materials. Surprisingly, while altermagnetism survives at large Hubbard $U$ in the system with anisotropy, the $d$-wave pairing shows an non-trivial enhancement when the $t_A$ increases, resulting in a dome-shape region in Fig.~\ref{fig1}(b), in which we measure the total vertex pairing function that sums over the momentum space. 


\begin{figure}[t!]
	\centering
    \includegraphics[width=\columnwidth]{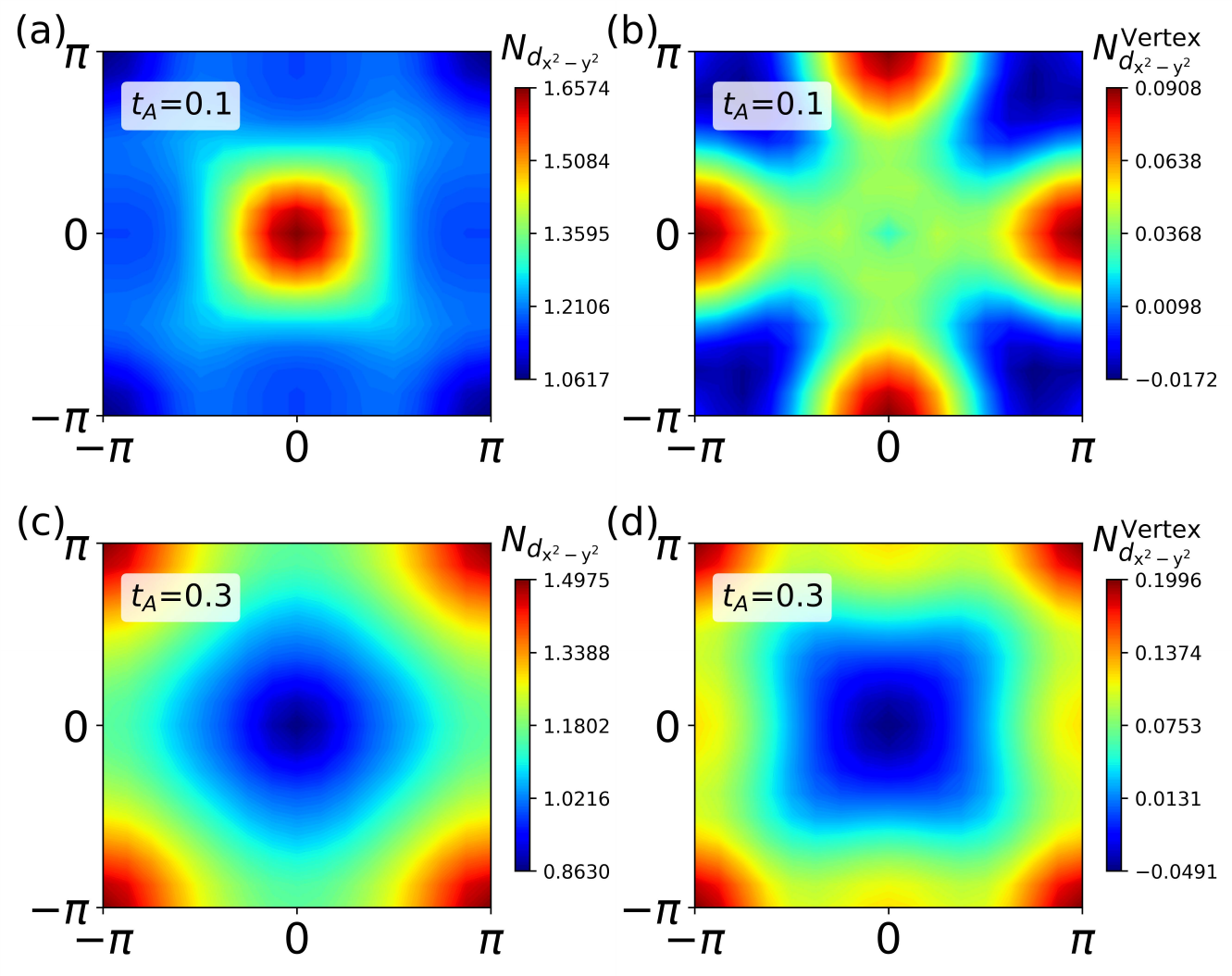}
    \caption{(Color online) Pairing mechanism changes with the anisotropy. (Left) The full and (Right) the vertex $d$-wave pairing function in the momentum space. With increasing $t_A$, the full pairing peak shifts from (0,0) to $(\pi,\pi)$, while the vertex pairing peak shifts from $(0,\pi)$ to $(\pi,\pi)$.
    }\label{fig4}
\end{figure}

The interplay of interaction and anisotropy could lead to the enhancement of unconventional pairings, such as $d$-wave and $p$-wave. The total vertex pairing function plotted in Fig.~\ref{fig3} shows the influence of the interaction on the pairing. The positive value indicates the  enhancement of the pairing, and the negative for the suppression of the pairing. Except the pairing of $d$-wave and $p_{\uparrow,\downarrow}$-wave, other channels show suppression in total effective pairing in the presence of repulsive Hubbard $U$. The results are consistent with the expectation that the repulsion generally suppresses the conventional $s$-wave pairing. We also find that the $d$-wave pairing is the most enhanced pairing channel among all in the presence of both anisotropy and repulsion. Compared to $U=1,2$, with strong $U=4$, the enhancement of effective $d$-wave pairing is significant against anisotropy. At high anisotropy $t_A\geq0.6$, the slight decay or flattening of the total vertex pairing function results from the ineffective pairing between spins in the extreme anisotropic surface of different spin species.

We show the $d$-wave pairing function in momentum space in Fig.~\ref{fig4}, in both full form and vertex form. The result shows that there is a drastic change of the full pairing peak from (0,0) to ($\pi$,$\pi$), when $t_A$ increase across 0.2. The corresponding change of maximum effective enhancement is also found, from (0,$\pi$) to ($\pi$,$\pi$). This change coincides with the drastic changing point of momentum domain spin polarization in Fig.~\ref{fig1}(a), signaling the presence altermagnetism will significantly affect the dominant pairing mechanism. The system with the momentum-dependent spin splitting eventually results in different preferences of the spin constituent of the pairs. At small $t_A$, approximate nesting of the Fermi surface still triggers the instability $E_{-k+Q}=E_{k}$ of unconventional pairing in the particle-particle channel, resulting in the peak at $\Gamma$ point, though the dominant instability is $E_{k+Q}=-E_{k}$ of antiferromagnetic order in the particle-hole channel. The enhancement of vertex pairing in (0,$\pi$) could be attributed by the Van Hove singularity at (0,$\pi$) in the absence of anisotropy. At larger $t_A$, the polarized spin patches near the border of momentum space become the main constitutes of the pairing, resulting in ($\pi$,$\pi$) peak in both full and vertex pairing. A recent functional renormalization group study also considers similar model with more parameters at finite temperature~\cite{Parthenios2025-ak}, by extrapolating the result to the low temperature, they shows the existence of the $d$-wave instability in some parameter range. This further augments our finding that the enhancement of the $d$-wave pairing by the anisotropy and interaction, in which the behind mechanism we would suggest in the follows.

\begin{figure}[t!]
	\centering
    \includegraphics[width=\columnwidth]{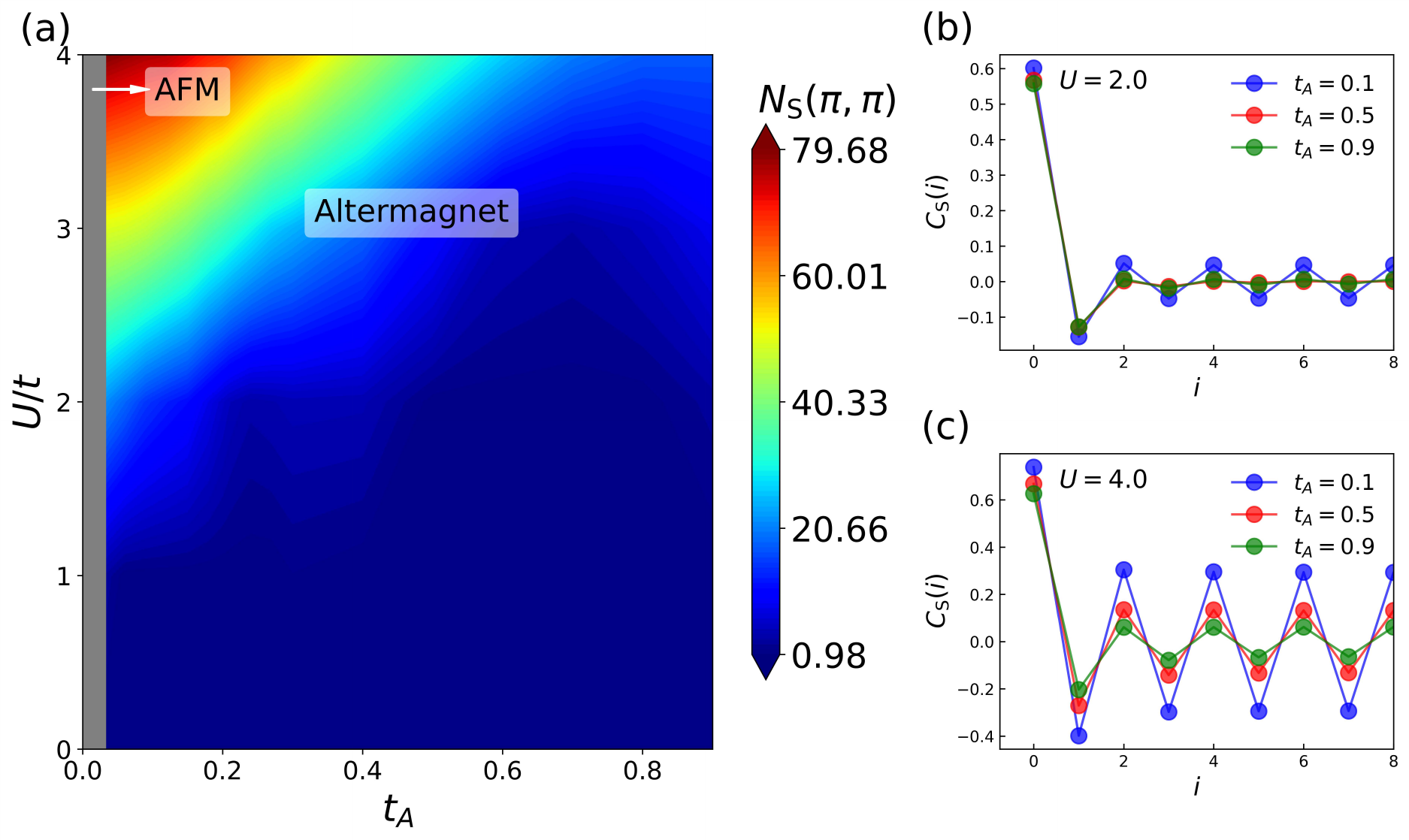}
    \caption{(Color online) Spin density wave measurement of varying $U/t$ and $t_A$. (a) The antiferromagnetic structure factor \(N_{\mathrm{S}}(\pi,\pi)\) with the magnitude represented by the color. The repulsion $U/t$ enhances the antiferromagnetic fluctuation while the anisotropy $t_A$ suppresses it. We also show the real-space spin correlation of (b) $U=2$ and (c) $U=4$ at various $t_A$, in which the antiferromagnetic signal is suppressed by $t_A$.
    }\label{fig5}
\end{figure}
\emph{Altermagnetic version of RVB theory}--The arise of the $d$-wave enhancement could be attributed to the suppression of other orders. The most prominent ordering in the system is the antiferromagnetic order, in which the instability is mainly triggered by Hubbard repulsion and augmented by the perfect nesting condition. We observe here how the spin density factor $(\pi,\pi)$ that can signal antiferromagnetism changes with the anisotropy that breaks the perfect nesting. In Fig.~\ref{fig5}, we observe suppression of $N_{\mathrm{S}}(\pi,\pi)$ by anisotropy, as well as significant enhancement by the Hubbard interaction. Compared with Fig.~\ref{fig1}(a), the region in which the $d$-wave pairing strongly enhanced matched the region that $N_{\mathrm{S}}(\pi,\pi)$ significantly suppressed.

The coincidence of suppressed antiferromagnetic order and the enhanced $d$-wave pairing links us to the renowned resonant valence bonding theory (RVB) that might provide explanation to high $T_c$ superconductivity. RVB states exist in the strongly correlated regime, where double occupancies is not allowed, and doping are required to introduce holes in the antiferromagent that resulted from the prefect nesting. In our model here, we are not changing the doping from half filling, but we suppress the antiferromagnetic order by means of altermagnetic type of anisotropic Fermi surface. Though no usual type of hole exists in our system, the momentum-dependent spin splitting provides a effectively hole region of different spin species. 

To unveil the mechanism of the enhancement, we examine the $t-J$ model at half filling with spin-dependent $d$-wave anisotropy hopping, in which we deduce the grand potential with $d$-wave pairing ($\Delta_{d}(k)=\Delta_{d}$), $\Omega = \sum_{\mathbf{k}} (\epsilon_{\mathbf{k}} -\left( E_{\mathbf{k}}^+ + E_{\mathbf{k}}^- \right)/2) + \Delta^2/J$, where $\epsilon_{\mathbf{k}}$ is the kinetic energy of fermion, $E_{\mathbf{k}}^{\pm}= \Bigl|\sqrt{\epsilon_{\mathbf{k}}^2+\Delta_{\mathbf{k}}^2}\pm|\epsilon_A|\Bigr|$ is the eigenvalues of the Hamiltonian. We find that at large $J$  the $d$-wave pairing state has the lower free energy than the $s$-wave pairing state even with the anisotropy $t_A$. This supports the possible scenario of arising $d$-wave pairing from suppression of antiferromagnetic order by breaking the Fermi surface nesting. The derivation is given in the supplementary materials.

\emph{Conclusions and outlook}-- In summary, we have shown that a spin-anisotropic Hubbard model near half-filling hosts a robust altermagnetic phase in the presence of on-site repulsion, with spin-momentum locking emerging from band anisotropy. Using quantum Monte Carlo simulations, we reveal that while the Hubbard interaction enhances antiferromagnetic correlations in the isotropic limit, increasing spin-dependent anisotropy suppresses magnetic order and instead favors the formation of $d$-wave pairing correlations. This enhancement is strongest where the antiferromagnetic structure factor is diminished, providing a striking parallel to the RVB mechanism, but realized without doping. Our results highlight altermagnetism as a fertile ground for unconventional superconductivity, driven by short-range spin fluctuations in the absence of perfect nesting. 

The other enhanced channel of $p$-wave pairing by the anisotropy is worth further investigation. Possible Larkin–Ovchinnikovj states could be resulted due to the spin-dependent anisotropy state. Also, how the pairing mechanism changes with doping in altermagnet is another interesting direction, as both doping and the anisotropy would destroy the nesting condition. This could probably give hints to the relation of the finding to doped cuprate, in which a recent studies shows magnetized oxygen can result in AM~\cite{Li2024-op}. Also, an interesting question of whether the pseduogap like regime could be resulted from the induction of d-wave pairing in altermagnetic deispersion is worth in-depth studies, it might hints the ultimate fomation mechanism of the cuprates' pseudogap. Overall, these findings open new possibilities for exploring correlated superconductivity in systems with engineered spin anisotropy, including ultracold atoms and synthetic magnetic platforms.

\emph{Acknowledgment.}-- This work is supported by the National Natural Science Foundation of China~(Grant No. 12204130), Shenzhen Key Laboratory of Advanced Functional CarbonMaterials Research and Comprehensive Application (Grant No. ZDSYS20220527171407017). 

\bibliography{ref}
\end{document}